\documentclass[twocolumn,english,aps,prl,floatfix,amssymb,superscriptaddress]{revtex4}
\usepackage[T1]{fontenc}
\usepackage[latin9]{luainputenc}
\usepackage{color}
\usepackage{amsmath}
\usepackage{amssymb}
\usepackage{graphicx}

\makeatletter
\@ifundefined{textcolor}{}
{%
 \definecolor{BLACK}{gray}{0}
 \definecolor{WHITE}{gray}{1}
 \definecolor{RED}{rgb}{1,0,0}
 \definecolor{GREEN}{rgb}{0,1,0}
 \definecolor{BLUE}{rgb}{0,0,1}
 \definecolor{CYAN}{cmyk}{1,0,0,0}
 \definecolor{MAGENTA}{cmyk}{0,1,0,0}
 \definecolor{YELLOW}{cmyk}{0,0,1,0}
}

\usepackage{babel}
\usepackage[bookmarks=false,linkcolor=blue,urlcolor=blue,colorlinks,citecolor=blue]{hyperref}
\@ifundefined{showcaptionsetup}{}{%
 \PassOptionsToPackage{caption=false}{subfig}}
\usepackage{subfig}
\makeatother

\usepackage{babel}
\begin{document}

\title{Spin liquids from Majorana Zero Modes in a Cooper Box }

\author{Eran Sagi $^{1}$, Hiromi Ebisu$^{1}$, Yukio Tanaka$^{2}$, Ady
Stern$^{1}$, and Yuval Oreg}

\affiliation{~Department of Condensed Matter Physics, Weizmann Institute of Science,
Rehovot, Israel 76100\\
 $^{2}$~Department of Applied Physics, Nagoya University, Nagoya
464-8603, Japan }
\begin{abstract}
We propose a path for constructing diverse interacting spin systems
from topological nanowires in Cooper Boxes. The wires are grouped
into a three-wire building block called an 'hexon', consisting of
six Majorana zero modes. In the presence of a strong charging energy,
the hexon becomes a Cooper box equivalent to two spin-$1/2$ degrees
of freedom. By considering arrays of hexons and controlling the distances
between the various wires, one can tune the Hamiltonian governing
the low-energy spins, thus providing a route for controllably constructing
interacting spin systems in one- and two-dimensions. We explicitly
present realizations of the one-dimensional spin-$1/2$ XXZ chain,
as well as the transverse field Ising model. We propose an experiment
capable of revealing the nature of critical points in such effective
spin systems by applying a local gate voltage and measuring the induced
charge at a distance. To demonstrate the applicability of this approach
to two-dimensions, we provide a scheme for realizing the topologically
ordered Yao-Kivelson spin-liquid model, which has a collective Majorana
edge mode, similar to the B-phase of Kitaev's honeycomb model. 
\end{abstract}
\maketitle
{\it Introduction:}  Quantum spin models are of paramount importance
in condensed matter physics. While spin-models were traditionally
devised to study magnetically ordered materials, they are nowadays
known to exhibit highly non-trivial behavior, such \textcolor{black}{as
diverse critical phenomena and topolo}gical order (see, e.g., \cite{Belavin1984,Kitaev2006,Yao2007}). 

An important mathematical tool used to uncover these non-trivial properties
is the fermionization of the spins to Majorana degrees of freedom,
which in a few notable cases leads to exact solutions. Important examples
are the Jordan-Wigner transformation \cite{Jordan1928}, which allows
for exact solutions of one-dimensional (1D) spin-$1/2$ models, such
as the XXZ and Ising models \cite{Lieb1961,Shastry1997,Wang1992},
as well as two-dimensional (2D) ones, such as Yao-Kivelson (YK) model
\cite{Yao2007}. A more recent example is the Kitaev transformation
\cite{Kitaev2006}, originally used to solve the Kitaev honeycomb
model and demonstrate the emergence of non-Abelian spin-liquid behavior. 

Recent strong evidence indicate the emergence of Majorana zero modes
(MZMs) on the edges of semiconductor nanowires with spin-orbit coupling,
which are in proximity to an $s$-wave superconductor \cite{Sau2010,Lutchyn2010,Oreg2010,Mourik2012,Rokhinson2012,Deng2012,Churchill2013,Das2012,Finck2013,Albrecht2016,Lutchyn2017}.
When a few such MZMs are placed in a quantum dot with strong Coulomb
interactions, a so called Majorana-Cooper-Box, or MZM island, is formed.
The MZMs in the island can be mapped onto spin degrees of freedom.
For example, considering four MZMs in an islands, each pair forms
a fermion, thus generating four degenerate states. Including the constraint
on the total number of particles in the box due to the strong Coulomb
interactions, an effective two-level system\textendash or a spin $1/2$\textendash is
formed. Indeed, similar ideas have been used to study the so-called
\textquotedbl{}topological Kondo effect\textquotedbl{} \cite{Beri2012,Altland2013,Beri2013,Zazunov2014,Altland2014,Plugge2016a},
and realize few spin-liquid models \cite{Barkeshli2015,Landau2016,Plugge2016}.
Such models are of interest due to their promise as platforms for
fault tolerant quantum computing.

The approach of constructing effective spin systems from MZMs is reciprocal
to the common fermionization of spin models: instead of starting with
physical spins and mapping them to Majorana degrees of freedom through
mathematical transformations, we begin with physical MZMs and map
them onto spins. In some cases, the resulting spin models may then
be solved through a distinct transformation to fictitious Majorana
degrees of freedom, which are non-local with respect to the physical
MZMs.

\begin{figure*}
\subfloat[\label{fig:hexon}]{\includegraphics[scale=0.53]{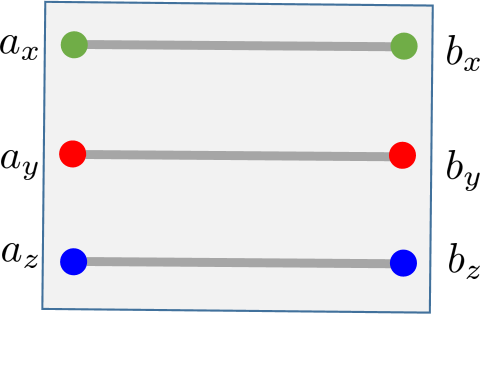}

}\subfloat[\label{fig:spinchain}]{\includegraphics[scale=0.53]{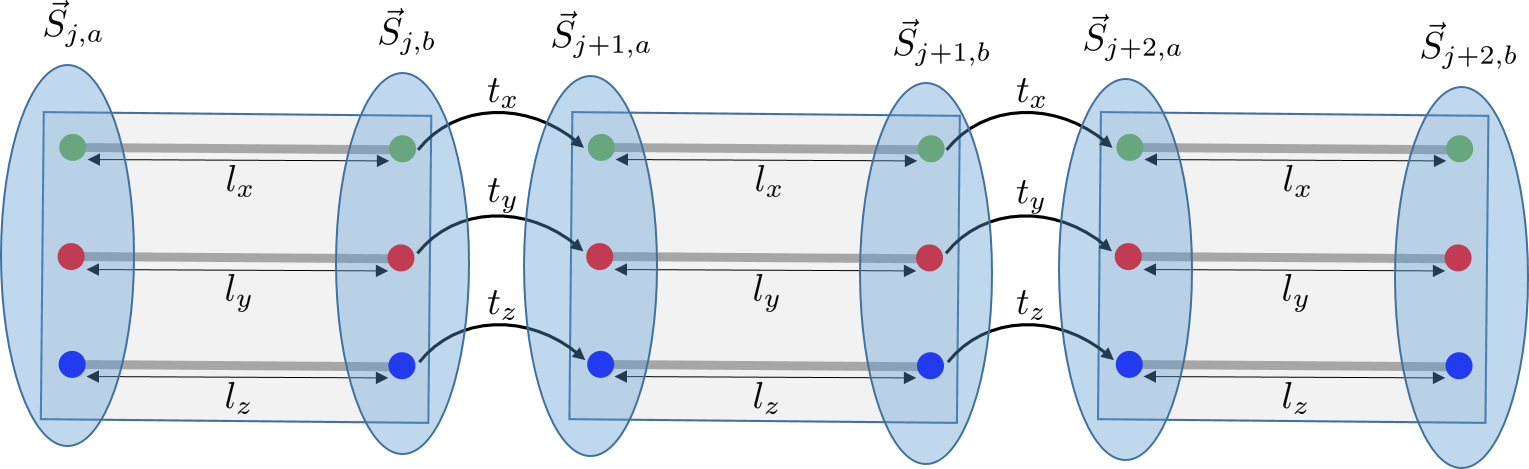}}

\subfloat[\label{fig: decoration building block}]{\includegraphics[scale=0.25]{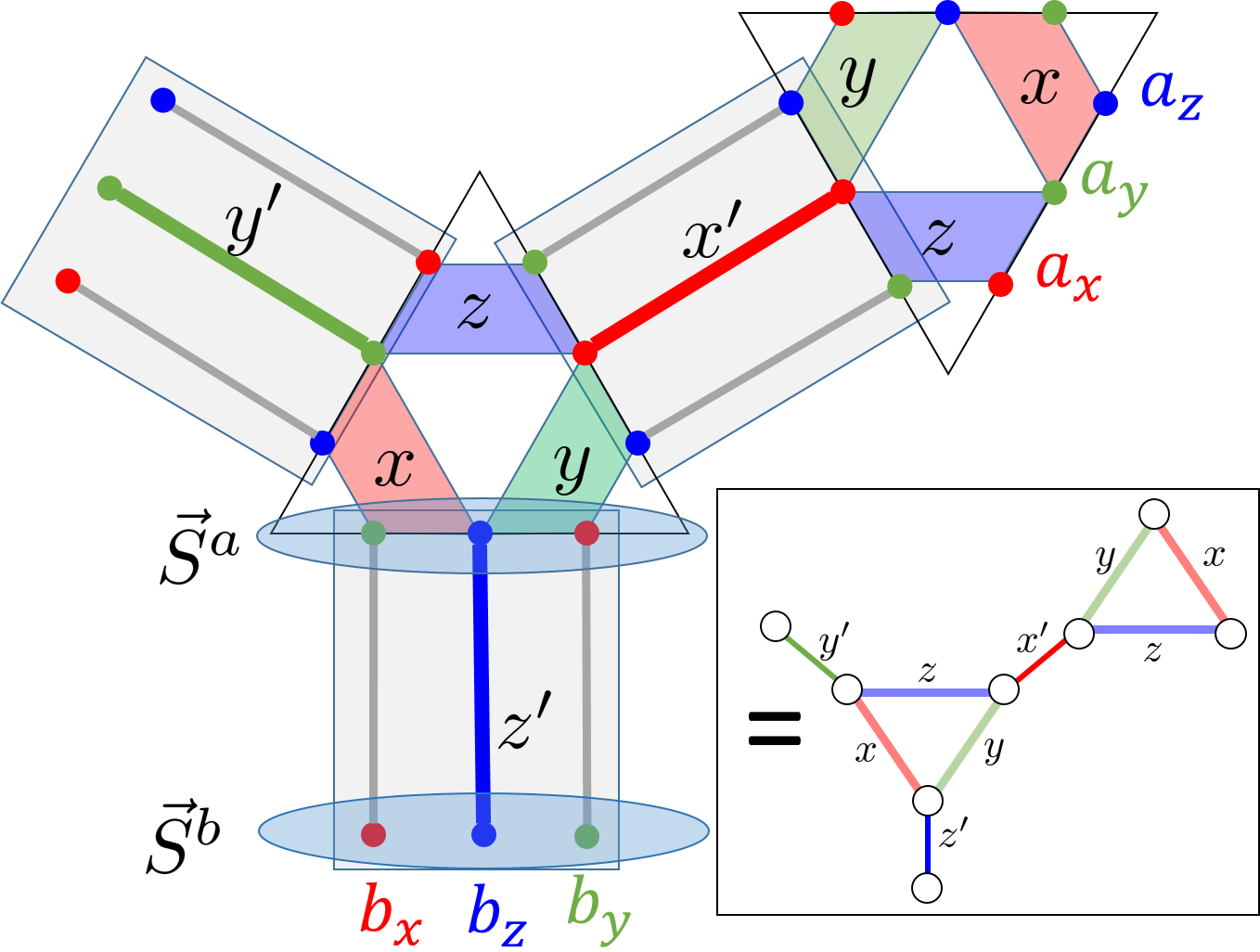}

}\subfloat[\label{fig:YK model}]{\includegraphics[scale=0.23]{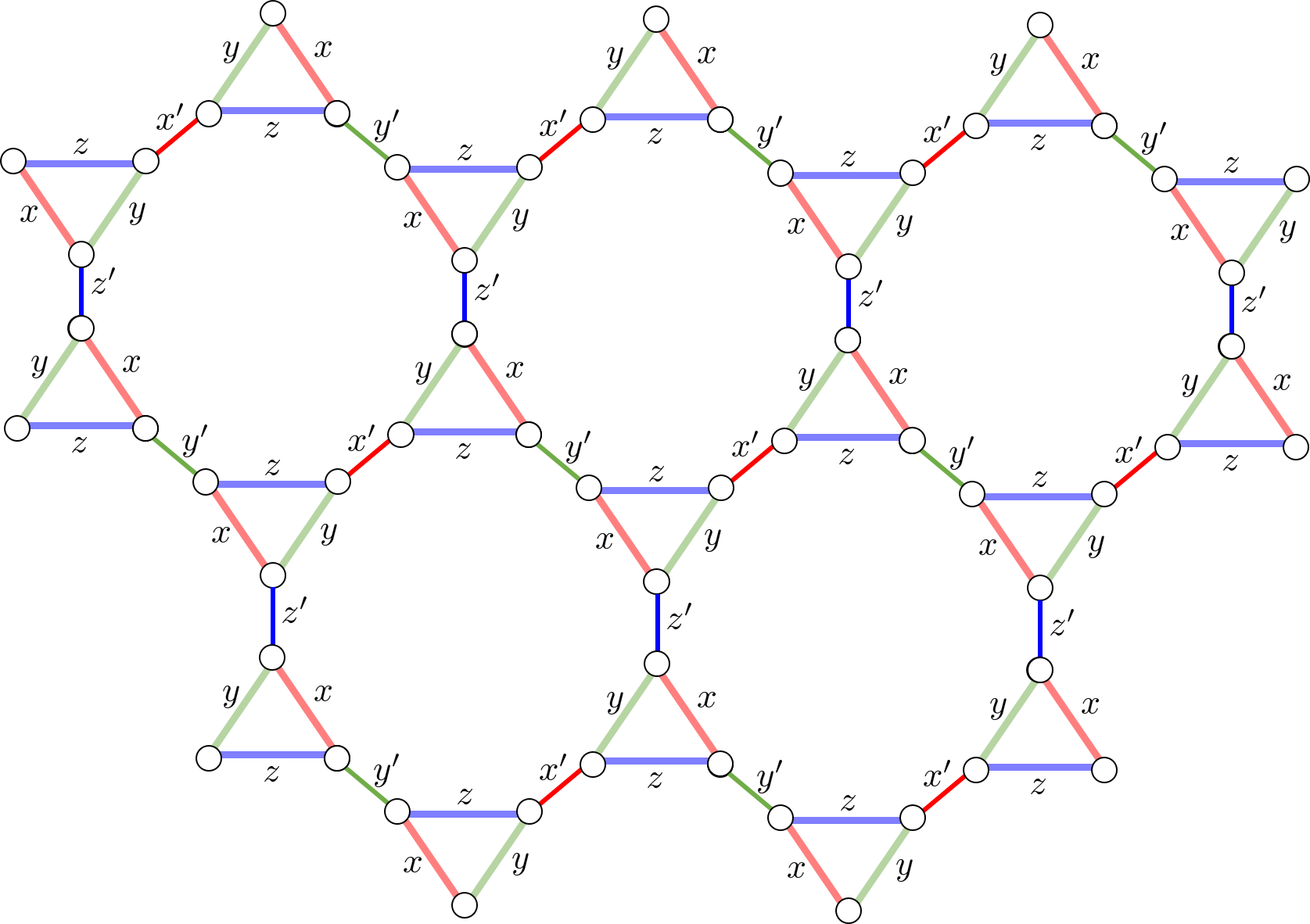}}

\caption{(a) The basic hexon building block, consisting of three semiconductor
nanowires. The wires are tuned to the topological regime, in which
each edge has a protected Majorana zero mode. In the presence of a
strong charging energy, we demonstrate that each hexon is equivalent
to two spin-$1/2$ degrees of freedom. (b) By forming an array of
hexons, we effectively model spin chains. (c) The hexons can also
be arranged in a 2D structure, giving rise to the Yao-Kivelson model
shown in Fig. (d).}
\end{figure*}

In this work we propose a different setup, where each box is made
of three semiconducting wires, as shown in Fig. \ref{fig:hexon},
and demonstrate that in the presence of a strong charging energy,
two effective spin degrees of freedom emerge at low energies. Due
to the presence of six MZMs, we refer to our building block as an
'hexon' \cite{Karzig2017}. The hexon building blocks are shown to
be highly tunable, and in fact, controlling the coupling between different
MZMs (e.g., by tuning the local chemical potential) allows us to fully
determine the coupling between different spins and the effective magnetic
field they experience. If many such building blocks are arranged in
a 1D line, or cover the 2D plane, this allows us to simulate a plethora
of spin models in 1D and 2D using only tunneling and local charging
terms. 

We start by focusing on the 1D setup shown in Fig. \ref{fig:spinchain},
and demonstrate that by controlling the couplings between different
MZMs, one can simulate $SU(2)$ invariant spin-$1/2$ chains. In particular,
we realize the spin-$1/2$ Heisenberg chain, known to be described
by a low-energy Luttinger Liquid (LL) fixed point. By modulating the
distance between a specific pair of MZMs as a function of time and
measuring the induced charge at a distance, we propose measurable
imprints of this critical point. 

We then provide a recipe for constructing the transverse field Ising
model, known to give rise to the Ising critical point. We propose
imprints of this critical point, which are in particular capable of
directly probing the properties of the so called $\sigma$-operator
\cite{DiFrancesco1997}. 

Finally, we describe the construction of the 2D YK decorated honeycomb
model (shown in Figs. \ref{fig: decoration building block}-\ref{fig:YK model}),
giving rise to a spin-liquid state with a chiral Ising CFT on the
edge, and discuss the experimental consequences of this gapless edge. 

{\it The hexon:}  The basic building block in our construction is
the so-called hexon \cite{Karzig2017}, illustrated in Fig. \ref{fig:hexon}.
Each hexon is composed of three semiconductor nanowires with strong
spin-orbit coupling. The wires are proximity coupled to an $s$-wave
superconductor. Applying a strong Zeeman field drives the wires into
the topological regime, in which protected MZMs reside near the ends
of each wire \cite{Sau2010,Lutchyn2010,Oreg2010,Mourik2012,Rokhinson2012,Deng2012,Churchill2013,Das2012,Finck2013,Albrecht2016}.
The Majorana zero modes are conveniently denoted by the operators
$a_{\alpha},b_{\alpha}$, with the indices $\alpha=x,y,z$, as illustrated
in Fig. \ref{fig:hexon}. The presence of six Majorana zero modes
leads to a degeneracy of 8.

Taking Coulomb blockade into account, and assuming that the charging
energy $E_{C}$ is the largest energy scale in play, we fix the charge,
and therefore the parity, of the entire hexon by controlling a back-gate
voltage. At low energies, the parity of the entire hexon can be written
in terms of the MZMs as
\begin{equation}
{\cal P}=ia_{1}a_{2}a_{3}b_{1}b_{2}b_{3}.\label{eq:constraint}
\end{equation}
Thus, by controlling the back-gate voltage, we can effectively apply
the constraint ${\cal P}=1$ (or similarly, ${\cal P}=-1$), thus
reducing the ground state degeneracy to 4. 

To find a useful parametrization of the remaining 4-dimensional low
energy subspace, we define spin-$1/2$ operators according to\textbf{
}\cite{Romito2017}
\begin{equation}
S_{a}^{x}=ia_{y}a_{z},\hphantom{cc}S_{a}^{y}=ia_{x}a_{z},\hphantom{cc}S_{a}^{z}=ia_{x}a_{y}\label{eq:Spina}
\end{equation}
\begin{equation}
S_{b}^{x}=ib_{y}b_{z},\hphantom{cc}S_{b}^{y}=ib_{x}b_{z},\hphantom{cc}S_{b}^{z}=ib_{x}b_{y}.\label{Spinb}
\end{equation}
It can easily be checked that these are in fact spin-$1/2$ operators
(i.e., they satisfy the relation $S^{l}S^{m}=i\epsilon_{klm}S^{k}+\delta_{lm}$),
which commute with the total parity {[}Eq. (\ref{eq:constraint}){]},
and therefore do not violate the parity-fixing constraint. The number
of states indeed coincides with the degeneracy of a two-spin system,
and we find that at low energies the six MZMs are reduced to \emph{two}
effective spin-1/2 degrees of freedom.  In what follows, the effective
spin degrees of freedom will be used to design non-trivial spin models
by engineering the coupling between different MZMs. 

{\it Coupling the spins:}  We start by studying the terms that arise
from coupling the MZMs within the hexon. The first of these arises
when the lengths $l_{\alpha}$ of the wires (see Fig. \ref{fig:spinchain})
are made short enough such that the Majorana wavefunctions at the
two ends overlap. In this case we get terms of the form 
\begin{equation}
H_{1}=i\sum_{\alpha=x,y,z}J_{\alpha}a_{\alpha}b_{\alpha},\label{eq:tunneling short wires-1}
\end{equation}
where the coupling constants $J_{\alpha}$ are controlled by the lengths
$l_{\alpha}$. Notice that the sign of $J_{\alpha}$ can also be tuned
as the overlap between the MZM wave-functions generically changes
sign as a function of $l_{\alpha}$. Alternatively, by tuning the
chemical potential in the wire, one controls the localization length
of the MZMs and therefore their coupling. Taking the constraint ${\cal P}=1$
into account, and using Eqs. (\ref{eq:Spina}) and (\ref{Spinb}),
we can write these as 
\begin{equation}
H_{1}=\sum_{\alpha=x,y,z}J_{\alpha}S_{a}^{\alpha}S_{b}^{\alpha}.\label{eq:spin spin-1}
\end{equation}

One can also generate a different set of terms by coupling MZMs of
the same type ($a$ with $a$ and $b$ with $b$), e.g., by changing
the distance between wires. This generates coupling terms of the form
\begin{align}
H_{2} & =i\sum_{\alpha\alpha'}\tilde{t}_{\alpha\alpha'}\left(a_{\alpha}a_{\alpha'}+b_{\alpha}b_{\alpha'}\right).\label{eq:H same side coupling}
\end{align}
In terms of the spin operators, $H_{2}$ can be written as 
\begin{equation}
H_{2}=\sum_{\alpha}B_{\alpha}\left(S_{a}^{\alpha}+S_{b}^{\alpha}\right),\label{Zeeman H}
\end{equation}
with $B_{\alpha}\propto\epsilon_{\alpha\beta\gamma}\tilde{t}_{\beta\gamma}$. 

To recapitulate, we find that each hexon is equivalent to two spins
degrees of freedom, and that the effective coupling between the two
spins, as well as coupling to an external magnetic field, can be controlled
by tuning the coupling between the MZMs (for example, with gate potentials).
In what follows we use these hexon building blocks to form 1D and
2D interacting spin models. 

{\it Realizing $SU(2)$-invariant spin chains:}  Consider the array
of hexons depicted in Fig. \ref{fig:spinchain}. As we discussed above,
these are equivalent to an array of spins, labeled by $\vec{S}_{j,\gamma},$
where $j$ enumerates the different hexon unit cells and $\gamma=a,b$
differentiates between the two spins in each unit cell. 

We start by assuming that the distance between different wires is
large such that the effective Zeeman field {[}$B_{\alpha}$ in Eq.
(\ref{Zeeman H}){]} vanishes, yet the lengths $l_{\alpha}$ are small
enough to generate $H_{1}$-type terms of the form

\begin{equation}
H_{1}=\sum_{j}\sum_{\alpha=x,y,z}J_{\alpha}S_{j,a}^{\alpha}S_{j,b}^{\alpha},\label{eq:spin spin}
\end{equation}

Coupling terms of the form $S_{j,b}^{\alpha}S_{j+1,a}^{\alpha}$ can
additionally be generated by bringing different hexons close to each
other. This generates tunneling terms of the form 
\begin{equation}
H_{\text{tunneling}}=i\sum_{\alpha=x,y,z}\tilde{t}'{}_{\alpha}\sum_{j}b_{\alpha j}a_{\alpha j+1}.\label{Tunneling between wires}
\end{equation}
These terms, however, alter the parity of the hexons and therefore
do not commute with the constraint. Under our assumption that the
charging energy $E_{C}$ is the largest energy scale, the tunneling
terms in Eq. (\ref{Tunneling between wires}) thus scale down to zero.
Nevertheless, we can form combinations of these terms that commute
with the constraint. The lowest order terms generated in perturbation
theory take the form 
\begin{align}
H'_{\text{1}} & =\sum_{\alpha=x,y,z}J'_{\alpha}\sum_{j}S_{j,b}^{\alpha}S_{j+1,a}^{\alpha}\label{eq:Spin Spin 2}
\end{align}
where $J'_{\alpha}\propto\frac{\Pi_{\alpha'\neq\alpha}\tilde{t}'_{\alpha'}}{E_{C}}$. 

At low energies, our model is therefore given by a combination of
Eqs. (\ref{eq:spin spin}) and (\ref{eq:Spin Spin 2}). For simplicity,
we start by assuming that the system was tuned to be $SU(2)$-invariant,
i.e., $J_{\alpha}=J$, and $J'_{\alpha}=J'$. We further assume that
$J,J'>0$.

Clearly, if $J>J'$, we get a fully gapped dimerized phase, in which
the two spins corresponding to each hexon form a singlet state. In
the opposite regime where $J'>J$, the system is again in a dimerized
phase, now with adjacent spins originating from different hexons forming
singlet states. 

The two above phases are topologically distinct, with the second state
giving rise to a protected decoupled spin on each edge. As such, we
expect to find a critical point if we tune $J=J'$. Indeed, at this
point our model becomes the spin-$\frac{1}{2}$ Heisenberg model,
known to be dual to a 1D model of interacting fermions. The latter
is described by the Luttinger-liquid fixed point Hamiltonian
\begin{equation}
H_{LL}=\frac{v}{2\pi}\int dx\left[K\left(\partial_{x}\theta\right)^{2}+\frac{1}{K}\left(\partial_{x}\varphi\right)^{2}\right],\label{eq:LL}
\end{equation}
with the Luttinger parameter $K=\frac{1}{2}$ \cite{Giamarchi}, the
spin operator $S_{z}(x)=\frac{1}{\pi}\partial_{x}\varphi$, and $\left[\theta(x),\varphi(x')\right]=i\pi\Theta(x-x')$.
The Luttinger parameter can be varied if the $SU(2)$ symmetry is
broken to $U(1)$, i.e., if one of the components $J_{\alpha}$ is
not the same as the other two. Indeed, mutual capacitance terms generically
renormalize the Luttinger parameter \cite{Glazman1997}. Notice that
we neglected higher order tunneling terms as a renormalization group
analysis indicates they are irrelevant.

{\it Experimental signature:}  The above constitutes an example of
realizing a critical spin model from the physical MZMs. It is natural
to ask whether one can measure imprints of the gapless spin model
in the current realization. Such an imprint is required to distinguish
between gapless and gapped states, as well as between gapless states
described by different conformal field theories (CFTs). 

Clearly, given that the charge degrees of freedom are gapped, one
cannot use electronic transport measurements. A possible route is
then to use thermal conductance measurements instead. While such measurements
are possible, and were in fact used recently to detect imprints of
the non-Abelian nature of the quantum Hall plateau at filling $5/2$
\cite{Mross2017,Banerjee2017,Wang2017}, they are difficult in practice. 

Instead, we propose an alternative experiment in which a time-dependent
gate modulating the coupling between two specific MZMs is applied.
If we choose these to be $a_{xj_{0}}$ and $a_{yj_{0}}$ (or similarly
$b_{xj_{0}}$ and $b_{yj_{0}}$) in a specific unit cell, we obtain
a time-dependent Hamiltonian of the form $H_{\text{pert}}=f(t)S^{z}(x_{0})$,
where for simplicity we assume that $f(t)=V_{0}\cos\left(\omega t+\phi_{0}\right)$. 

To find imprints of the gapless nature of the underlying state, we
propose to measure the expectation value of $S_{j,a}^{z}=ia_{xj}a_{yj}$
(or similarly $S_{j,b}^{z}=ib_{xj}b_{yj})$ in a distant unit cell
$j$, at later times. We refer to this quantity as the \textit{induced
parity} of these MZMs. Performing linear response, the value of $S^{z}(x,t)$,
i.e., the induced parity of the appropriate pair of MZMs at point
$x$ and time $t$, is given by 
\begin{equation}
\left\langle S^{z}(x,t)\right\rangle =\int dt'f(t')\chi(t-t',x-x_{0}),\label{eq:S_z}
\end{equation}
with $\chi$ being the dynamic susceptibility: $\chi(t-t',x-x')=i\left\langle \left[S^{z}(x,t),S^{z}(x',t')\right]\right\rangle \Theta(t-t')$.
As we demonstrate in the Appendix, in a \textit{non-chiral} critical
point, where the (time-ordered) correlation function takes the form
$G\sim\alpha^{4h}/(x^{2}-v^{2}t^{2})^{2h}$ (with $\alpha$ being
the short distance cutoff, and $h$ the conformal dimension), we obtain
\begin{align}
\left\langle S^{z}(x,t)\right\rangle  & =\frac{V_{0}\alpha^{4h}}{v^{2h+\frac{1}{2}}}\left(\frac{\omega}{\left|\Delta x\right|}\right)^{2h-\frac{1}{2}}\nonumber \\
 & \times\Re\left\{ Be^{i\left(\omega t+\phi_{0}\right)}K_{\frac{1}{2}-2h}\left(i\frac{\omega\left|\Delta x\right|}{v}\right)\right\} \label{eq:Sz}
\end{align}
where $\Delta x=x-x_{0},$ $B$ is a complex dimensionless constant,
and $K_{n}(z)$ is the modified Bessel function of the second kind.
Using the asymptotic form $K_{n}(z)\sim\sqrt{\frac{\pi}{2z}}e^{-z}$,
we obtain $\left\langle S^{z}(x,t)\right\rangle \propto\frac{V_{0}\alpha^{4h}}{v^{2h}}\frac{\omega^{2h-1}}{\left|\Delta x\right|^{2h}}\cos\left[\omega\left(t-\frac{\left|\Delta x\right|}{v}\right)+\varphi_{0}\right]$
at large distances, where $\varphi_{0}$ is a constant phase.\textcolor{red}{{}
}By measuring the induced parity at a distance from the perturbation,
we can thus get an imprint of the critical nature of the transition,
and in particular, directly measure the critical exponent $h$. Moving
away from the critical point, Eq. (\ref{eq:Sz}) becomes an exponential
decay. 

In our case, the Luttinger-liquid description in Eq. (\ref{eq:LL})
implies that the $S_{z}-S_{z}$ time ordered correlation function
is described by $h=1/2$, leading to $\left\langle S^{z}(x,t)\right\rangle \propto\frac{V_{0}\alpha^{2}}{v\left|\Delta x\right|}\cos\left[\omega\left(t-\frac{\left|\Delta x\right|}{v}\right)+\phi_{0}\right]$
(see the Appendix for more details). 

{\it The transverse field Ising model:}  The flexibility of altering
the various length scale in our setup allows us to realize a large
set of spin models which goes beyond the above $SU(2)$ invariant
chains. In what follows we provide an explicit construction of another
prominent spin chain - the transverse field Ising model - defined
by the Hamiltonian

\begin{equation}
H_{\text{Ising}}=\sum_{j}\left[-JS_{j}^{z}S_{j+1}^{z}+hS_{j}^{x}\right].\label{eq:ising}
\end{equation}

The first term can be generated similarly to the above: by making
the length $l_{z}$ of the $z$-type wires short enough and simultaneously\textcolor{red}{{}
}bringing $x$ and $y$ type wires coming from adjacent hexons closer
to each other. If these terms are taken to have identical amplitudes,
they generate the first term in Eq. (\ref{eq:ising}). In addition,
assuming the distance between the $y$- and $z$-wires in each hexon
is made short, we generate $H_{2}$-type terms, giving rise to the
second term in Eq. (\ref{eq:ising}).

As is well known, the transverse field Ising model possesses two different
phases. For $J>h$, the ground state of the system spontaneously breaks
the $S^{z}\rightarrow-S^{z}$ symmetry and the spins collectively
point in the $\pm z$ direction. In the opposite regime, where $h>J$,
the state is not in a symmetry broken phase, and is connectable to
the state in which all the spins form an eigenstate of $S_{x}$ with
eigenvalue -1. The above two phases are separated by a gapless critical
point at $h=J$, in which case the effective spin chain is described
by an Ising fixed point with central charge $c=\frac{1}{2}$ \cite{BELAVIN1984333}. 

To probe this critical point, we can repeat the above experiment where
a local time dependent gate modulates the $z$ component of the magnetic
field at point $x_{0}$, and a charge probe at point $x$ effectively
measures the induced $S_{z}(x,t)$. Since $S_{z}$ can be identified
with the $\sigma$ primary field of the Ising CFT at low energies,
its correlation function scales with $h=1/16$. We can find the induced
parity by plugging this into Eq. (\ref{eq:Sz}). The dependence of
the parity on the distance and frequency provides a direct imprint
of the non-trivial CFT. 

{\it The 2D Yao-Kivelson spin liquid:}  In the above analysis, we
have demonstrated that the hexon building blocks provide a fruitful
playground for realizing 1D spin chains. As we argue now, the same
ideas can be applied to 2D spin models. To demonstrate this, we explicitly
construct the so-called Yao-Kivelson model \cite{Yao2007}, which
realizes a non-Abelian spin-liquid state. 

To do that, we sort the hexons in structures similar to Fig. \ref{fig: decoration building block}.
Notice that the labels $x,y,z$ of the MZMs are now alternating. In
each hexon, we assume that the colored wire is made short and therefore
induces $S_{a}^{\alpha}S_{b}^{\alpha}$-type terms. Correlated tunneling
terms between different hexons also generate $S_{a}^{\alpha}S_{b}^{\alpha}$-type
terms, with $\alpha$ determined by geometry - i.e., $\alpha$ is
chosen such that $\Pi_{\alpha'\neq\alpha}\tilde{t}'_{\alpha'}$ is
maximized. The resulting dominating terms are shown in Fig. \ref{fig: decoration building block}
in terms of the MZMs and in terms of the spin degrees of freedom in
the inset. 

If many such building blocks are connected in a way that covers the
2D plane, we obtain the decorated honeycomb lattice geometry, shown
in Fig \ref{fig:YK model}, where each link is given a label $\alpha$,
stating the dominating $S^{\alpha}S^{\alpha}$ term. The resulting
spin Hamiltonian is identical to the YK Hamiltonian, known to generate
a non-Abe\textcolor{black}{lian spin liquid state in the so-called
B-phase (as long as the coupling at the $x',y',$ and $z'$ links
is not too large), which in addition spontaneously breaks time reversal
symmetry. The Abelian A-phase of the Kitaev honeycomb model can also
be realized, for example, if we take the $z'$ coupling to be much
larger than $x',y'$. Other proposals for realizing this phase were
given in Refs. \cite{Plugge2016,Landau2016}. The advantage of the
current proposal is the ability to control all the coupling terms
with gate potentials.}

Within the B-phase, the edge of the sample gives rise to a chiral
Ising CFT, similar to the edge of a $p+ip$ superconducting state,
which can be constructed from arrays of Majorana wires as well \cite{Seroussi2014,Sagi2017}.
As opposed to the $p+ip$ superconducting state, however, the resulting
state is topologically ordered, with the $\sigma$-particle being
deconfined. \textcolor{black}{We note that one can obtain the above
spin-liquid from the $p+ip$ superconducting state by condensing $h/e$
vortices \cite{Wang2013}. }

In order to measure imprints of the gapless edge, we again repeat
the experiment above on the edge. As shown in the Appendix, the induced
$S_{z}$ is given by
\begin{equation}
\left\langle S^{z}(x,t)\right\rangle =V_{0}\frac{\omega^{2h-1}\alpha^{2h}}{v^{2h}}\cos\left[\omega\left(t-\frac{\Delta x}{v}\right)+\varphi_{0}\right],\label{eq:Sz-chiral}
\end{equation}
with $h$ being the smallest dimension among the operators excited
by $S^{z}$. In contrast to the 1D case, here the perturbation operates
on a chiral edge, and therefore cannot act as the $\sigma$ primary
field on that edge alone. 

{\it Conclusions:}  To summarize, in this manuscript we use MZMs
as a basic building block for constructing non-trivial spin models.
We show that a system containing six MZMs in a Majorana-Cooper box\textendash an
hexon\textendash is equivalent to two spin $1/2$ degrees of freedom.
By changing the coupling between the different MZMs, one can controllably
simulate spin models and tune them to criticality. We provided explicit
examples for the XXZ model and the transverse field Ising model in
1D, as well as the YK model in 2D. We have discussed possible physical
measurements capable of revealing the nature of these spin models
at critical points.

\textcolor{black}{We note that in 1D models, disorder can generally
have drastic effects, and in particular change the character of critical
points. Two-dimensional topologically ordered systems, such as the
YK spin liquid, are protected against weak disorder. In particular,
the gapless edges are generically protected by virtue of their chiral
nature. }
\begin{acknowledgments}
We acknowledge discussions with A. Altland, N. Andrei, R. Egger, A.
Kesselman, S. Kivelson, R. Lutchyn, and D. Pikulin. This work was
supported by the Deutsche Forschungsgemeinschaft (CRC 183), the Israel
Science Foundation (ISF), the Binational Science Foundation (BSF),
the European Research Council under the European Community's Seventh
Framework Program (FP7/2007-2013)/ERC Grant agreement No. 340210,
Microsoft Station Q, the Adams Fellowship Program of the Israel Academy
of Sciences and Humanities, a Grant-in-Aid for Scientific Research
on Innovative Areas Topological Material Science JPSJ KAKENHI (Grants
No. JP15H05851,No. JP15H05853, and No. JP15K21717), a Grant-in-Aid
for Scientific Research B (Grant No. JP15H03686 and JP18H01176), and
the JSPS overseas research fellowship.
\end{acknowledgments}

\bibliographystyle{apsrev4-1}
\bibliography{ref}

\newpage 
\onecolumngrid
\appendix
\section*{\large Appendix}

In this Appendix, we provide an explicit calculation of the response
to the oscillating magnetic field. We study the the response of chiral
and non-chiral one-dimensional critical systems. 

\section*{The Response Function}

In the main text, we proposed that the critical nature of our 1D spin
models can be revealed by studying the effect of perturbations of
the form $H_{\text{pert}}=f(t)S^{z}(x_{0})$, with $f(t)=V_{0}\cos\left(\omega t+\phi_{0}\right)$.
To do that, we would like to compute the average value of $S^{z}$
at a distant point $x$ and later times.

Performing linear response, this can be written as 
\begin{equation}
\left\langle S^{z}(x,t)\right\rangle =\int dt'f(t')\chi(t-t',x-x_{0}),\label{eq:S_z-1}
\end{equation}
with $\chi$ being the dynamic susceptibility: $\chi(t-t',x-x')=i\left\langle \left[S^{z}(x,t),S^{z}(x',t')\right]\right\rangle \Theta(t-t')$.
In our case, the function $f$ is harmonic, meaning we may write 
\begin{align}
\left\langle S^{z}(x,t)\right\rangle  & =V_{0}\Re\left\{ e^{i\phi_{0}}\int dt'e^{i\omega t'}\chi(t-t',x-x_{0})\right\} \nonumber \\
 & =V_{0}\Re\left\{ e^{i\omega t}e^{i\phi_{0}}\chi(\omega,x-x_{0})\right\} ,\label{eq:Sz in terms of FT}
\end{align}
where $\chi(\omega,x)=\int dte^{-i\omega t}\chi(t,x)$ is the frequency
domain form of the dynamic susceptibility.

The susceptibility can be written in terms of the time-ordered propagator
as 
\[
\chi(t,x)=-2\Theta(t)\text{Im}\left\{ G(t,x)\right\} .
\]
The time ordered propagators of one-dimensional CFTs can generally
be written as 
\[
G(t,x)=\frac{\alpha^{2(h+\bar{h})}}{\left[x-vt+i\epsilon\text{sign}(t)\right]^{2h}\left[x+vt-i\epsilon\text{sign}(t)\right]^{2\bar{h}}},
\]
where $h,\bar{h}$ are the conformal dimensions of the corresponding
field. For a non-chiral field, we have $h=\bar{h}$. For a chiral
field, one of these vanishes. 

\subsection*{Non-chiral fields}

For non-chiral fields, the susceptibility is given by 
\begin{align*}
\chi(t,x) & =-2\Theta(t)\alpha^{4h}\text{Im}\left\{ \frac{1}{\left[x-vt+i\epsilon\right]^{2h}\left[x+vt-i\epsilon\right]^{2h}}\right\} \\
 & =-2\Theta(t)\alpha^{4h}\text{Im}\left\{ e^{-2h\log A}\right\} ,
\end{align*}
where we follow Ref. \cite{Giamarchi} in defining 
\begin{align*}
A & =\left(x-vt+i\epsilon\right)\left(x+vt-i\epsilon\right)\\
 & =x^{2}-v^{2}t^{2}+2i\epsilon t.
\end{align*}
If we put the branch cut of the log along the negative real axis,
we get an imaginary part only for $x^{2}-v^{2}t^{2}<0,$ i.e. $t>\frac{\left|x\right|}{v},$
and we get 
\[
e^{-2h\log A}=e^{-2h\log\left|x^{2}-v^{2}t^{2}\right|-2\pi ih\Theta\left(t+\frac{x}{v}\right)\Theta\left(t-\frac{x}{v}\right)}.
\]
We therefore obtain 
\begin{equation}
\chi(t,x)=-2\sin\left(2\pi h\right)\alpha^{4h}\frac{\Theta(t)\Theta\left(t+\frac{x}{v}\right)\Theta\left(t-\frac{x}{v}\right)}{\left(v^{2}t^{2}-x^{2}\right)^{2h}}.\label{chi(t)}
\end{equation}

If $h=1/2$, this expression vanishes. This result is non-physical,
and indeed, the case $h=1/2$ requires special attention. In this
case, we can explicitly write 
\begin{align}
\chi(t,x) & =-2\Theta(t)\alpha^{2}\text{Im}\left\{ \left(\frac{1}{x-vt+i\epsilon}\right)\left(\frac{1}{x+vt-i\epsilon}\right)\right\} \nonumber \\
 & =-2\Theta(t)\alpha^{2}\text{Im}\left\{ \left[\mathcal{P}\frac{1}{x-vt}-i\pi\delta\left(x-vt\right)\right]\left[\mathcal{P}\frac{1}{x+vt}+i\pi\delta\left(x+vt\right)\right]\right\} \nonumber \\
 & =-2\pi\Theta(t)\alpha^{2}\left[\frac{1}{x-vt}\delta\left(x+vt\right)-\frac{1}{x+vt}\delta\left(x-vt\right)\right],\label{eq:gamma=00003D1}
\end{align}
where $\mathcal{P}$ denotes the principal value. 

In order to evaluate Eq. \ref{eq:Sz in terms of FT} for a general
$h$, we wish to get the frequency domain form of $\chi(t,x)$ in
Eq. \ref{chi(t)}:
\begin{align*}
\chi(\omega,x) & =2\alpha^{4h}\sin\left(2\pi h\right)\int_{\frac{\left|x\right|}{v}}^{\infty}dt\frac{e^{-i\omega t}}{\left(v^{2}t^{2}-x^{2}\right)^{2h}}\\
 & =\frac{2\alpha^{4h}\sin\left(2\pi h\right)}{v\left|x\right|^{4h-1}}\int_{1}^{\infty}dT\frac{e^{-i\frac{\omega\left|x\right|}{v}T}}{\left(T^{2}-1\right)^{2h}},
\end{align*}
where we have defined $T=\frac{vt}{\left|x\right|}$. Performing the
integral, we obtain 
\begin{equation}
\chi(\omega,x)=B\alpha^{4h}v^{-2h-\frac{1}{2}}\left(\frac{\omega}{\left|x\right|}\right)^{2h-\frac{1}{2}}K_{\frac{1}{2}-2h}\left[i\frac{\omega\left|x\right|}{v}\right],\label{eq:final chi}
\end{equation}
with \textbf{$B=\frac{2\sin\left(2\pi h\right)\Gamma\left(1-2h\right)\left(-2i\right)^{\frac{1}{2}-2h}}{\sqrt{\pi}}$}.
Notice that while Eq. \ref{chi(t)} vanishes for $h=1/2$, Eq. \ref{eq:final chi}
has a finite limit for $h\rightarrow1/2$. In this case, since $\sin\left(2\pi h\right)\Gamma\left(1-2h\right)\rightarrow const$
as $h\rightarrow1/2$, and $K_{-\frac{1}{2}}\left[z\right]=\sqrt{\frac{\pi}{2}}\frac{e^{-z}}{\sqrt{z}}$,
we get 
\[
\chi(\omega,x)\propto\alpha^{2}\frac{e^{-i\frac{\omega\left|x\right|}{v}}}{v\left|x\right|}.
\]
 We can obtain this result directly from Eq. \ref{eq:gamma=00003D1}.
In this case 
\begin{align*}
\chi(\omega,x) & =-2\pi\alpha^{2}\int_{0}^{\infty}dt\left[\frac{1}{x-vt}\delta\left(x+vt\right)-\frac{1}{x+vt}\delta\left(x-vt\right)\right]e^{-i\omega t}\\
 & =\frac{\pi\alpha^{2}}{v}\left[-\frac{1}{x}\Theta(-x)e^{i\omega\frac{x}{v}}+\frac{1}{x}\Theta(x)e^{-i\omega\frac{x}{v}}\right]=\frac{\pi\alpha^{2}}{v}\frac{e^{-i\omega\frac{\left|x\right|}{v}}}{\left|x\right|}.
\end{align*}

\subsection*{Chiral fields }

For chiral (right moving) fields, the susceptibility is given by 
\begin{align*}
\chi(t,x) & =-2\alpha^{2h}\Theta(t)\text{Im}\left\{ \frac{1}{\left[x-vt+i\epsilon\right]^{2h}}\right\} \\
 & =-2\alpha^{2h}\Theta(t)\text{Im}\left\{ e^{-2h\log A}\right\} ,
\end{align*}
where now 
\begin{align*}
A & =x-vt+i\epsilon.
\end{align*}
 The same analysis as in the non-chiral case indicates that 
\[
e^{-2h\log A}=e^{-2h\log\left|x-vt\right|-2\pi ih\Theta\left(t-\frac{x}{v}\right)},
\]
 and therefore 
\begin{equation}
\chi(t,x)=-2\sin\left(2\pi h\right)\alpha^{2h}\frac{\Theta(t)\Theta\left(t-\frac{x}{v}\right)}{\left(vt-x\right)^{2h}}.\label{chi(t)-1}
\end{equation}
Calculating the Fourier transform, we obtain 
\begin{align*}
\chi(\omega,x) & =\frac{2\alpha^{2h}\sin\left(2\pi h\right)}{v\cdot x^{2h-1}}\int_{1}^{\infty}dT\frac{e^{-i\frac{\omega x}{v}T}}{\left(T-1\right)^{2h}}\\
 & =C\frac{\omega^{2h-1}\alpha^{2h}}{v^{2h}}e^{-i\frac{\omega x}{v}},
\end{align*}
with $C=2\sin\left(2\pi h\right)i^{2h-1}\Gamma(1-2h)$.
\end{document}